\documentclass[sigconf]{acmart}

\AtBeginDocument{%
  }

\copyrightyear{2025}
\acmYear{2025}
\setcopyright{acmlicensed}\acmConference[CHI '25]{CHI Conference on Human Factors in Computing Systems}{April 26-May 1, 2025}{Yokohama, Japan}
\acmBooktitle{CHI Conference on Human Factors in Computing Systems (CHI '25), April 26-May 1, 2025, Yokohama, Japan}
\acmDOI{10.1145/3706598.3713881}
\acmISBN{979-8-4007-1394-1/25/04}

\usepackage{enumitem}
\usepackage{url}
\usepackage{pifont} 
\usepackage{tabularx}
\usepackage{multirow}
\usepackage{enumitem}
\usepackage{amsmath}
\usepackage{mathtools}
\usepackage{makecell}

\newcommand{\mathTextItalics}[1]{\text{\textit{#1}}}

\begin{document}

\title[GistVis: Automatic Generation of Word-scale Visualizations from Data-rich Documents]{GistVis: Automatic Generation of Word-scale Visualizations from Data-rich Documents}
\author{Ruishi Zou}
\authornote{Both authors contributed equally to this research.}
\affiliation{%
  \institution{Tongji University}
  \city{Shanghai}
  \country{China}}
\email{zouruishi@tongji.edu.cn}

\author{Yinqi Tang}
\authornotemark[1]
\affiliation{%
  \institution{Tongji University}
  \city{Shanghai}
  \country{China}}
\email{2054136@tongji.edu.cn}

\author{Jingzhu Chen}
\affiliation{%
  \institution{Tongji University}
  \city{Shanghai}
  \country{China}}
\email{2253543@tongji.edu.cn}

\author{Siyu Lu}
\affiliation{%
  \institution{Tongji University}
  \city{Shanghai}
  \country{China}}
\email{2250898@tongji.edu.cn}

\author{Yan Lu}
\affiliation{%
  \institution{Tongji University}
  \city{Shanghai}
  \country{China}}
\email{2253887@tongji.edu.cn}

\author{Yingfan Yang}
\affiliation{%
  \institution{Tongji University}
  \city{Shanghai}
  \country{China}}
\email{yangyingfan@tongji.edu.cn}

\author{Chen Ye}
\authornote{Chen Ye is the corresponding author.}
\affiliation{%
  \institution{Tongji University}
  \city{Shanghai}
  \country{China}}
\email{yechen@tongji.edu.cn}

\renewcommand{\shortauthors}{Zou et al.}

\begin{abstract}
Data-rich documents are ubiquitous in various applications, yet they often rely solely on textual descriptions to convey data insights. Prior research primarily focused on providing visualization-centric augmentation to data-rich documents. However, few have explored using automatically generated word-scale visualizations to enhance the document-centric reading process. As an exploratory step, we propose GistVis, an automatic pipeline that extracts and visualizes data insight from text descriptions. GistVis decomposes the generation process into four modules: Discoverer, Annotator, Extractor, and Visualizer, with the first three modules utilizing the capabilities of large language models and the fourth using visualization design knowledge. 
Technical evaluation including a comparative study on Discoverer and an ablation study on Annotator reveals decent performance of GistVis. Meanwhile, the user study (N=12) showed that GistVis could generate satisfactory word-scale visualizations, indicating its effectiveness in facilitating users' understanding of data-rich documents (+5.6\% accuracy) while significantly reducing their mental demand (p=0.016) and perceived effort (p=0.033).
\end{abstract}

\ccsdesc[500]{Human-centered computing~Information visualization}
\ccsdesc[500]{Human-centered computing~Visualization systems and tools}

\keywords{Word-scale visualization, Automatic visualization, Natural language processing, Interactive article, Data document}

\begin{teaserfigure}
    \centering
    \includegraphics[width=1\textwidth]{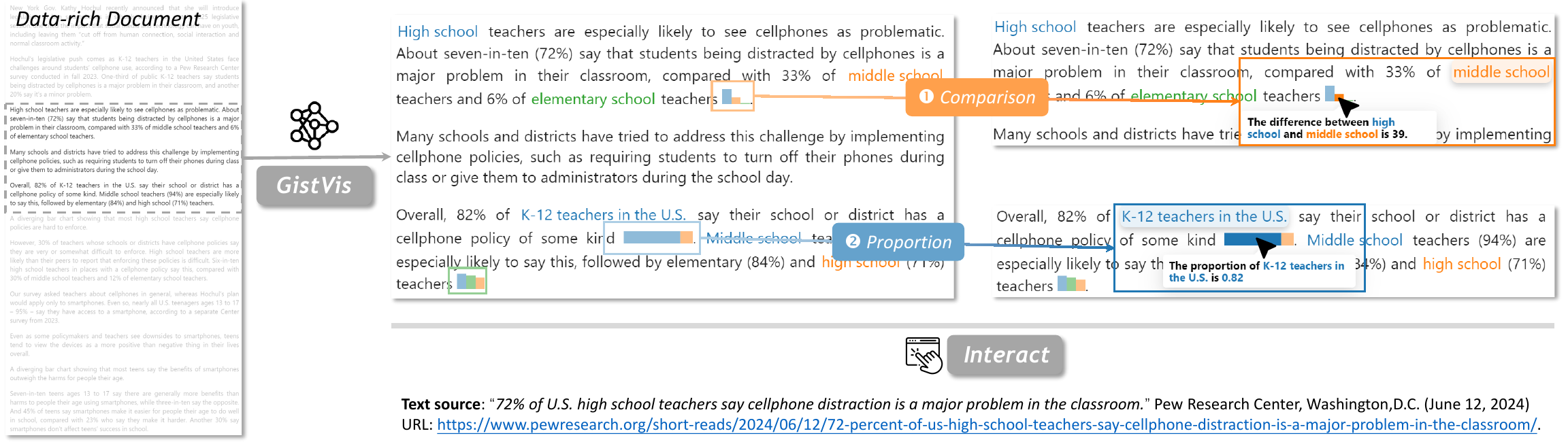}
    \caption{GistVis enables automatic generation of word-scale visualizations from data-rich documents. GistVis supports visualization generation for six data insight types: comparison \ding{182}, proportion \ding{183}, trend, rank, extreme, and value. GistVis also connects the word-scale visualizations and related entities in text snippets, allowing users to explore data-rich documents interactively.}
    \label{fig:teaser}
\end{teaserfigure}

\maketitle

\section{Introduction}
Data-rich documents, which use text descriptions to convey data insights, are prevalent in various applications, such as scientific papers~\cite{beck2017wordsized} and news articles~\cite{lin2018vizbywiki}.
Although researchers have long promoted using visualizations to convey data insights~\cite{washburne1927experimental, costigan-eaves1986edward}, authoring data-rich documents, especially pairing them with visualization, is time-consuming and requires extensive expertise~\cite{chen2022crossdata}. Such difficulty dictates that visualizations rarely exist in most data-rich documents.

Prior research proposed two directions to provide visualization for data-rich documents. One line of research proposed resorting to outside data sources to retrieve visualizations (e.g., \cite{lin2018vizbywiki, metoyer2018Coupling}) related to the text context. Although retrieved visualizations present relevant information to the text, the visualizations might not precisely match the text narrations. Others used the internal information within data-rich documents (such as tables and numbers) to generate visualizations~\cite{masson2023Charagraph, badam2019elastic, cui2020texttoviz}. Although using internal data sources ensures the visualization is pertinent to the document, most approaches use ``hard rules'' such as regular expressions to extract numerical data, ignoring the contextual data insights embedded within or between sentences.

Broadly, prior works have predominantly applied a visualization-centric approach~\cite{goffin2020Interaction} to augment existing data-rich documents. In a visualization-centric approach, visualizations are treated as independent constituents equal to or more important than the source document. In contrast, few works have explored an automatic method to generate visualizations for document-centric analysis of data-rich documents, where visualization only provides assistive contextual information to the original document. Motivated by this gap, we raised our research question: \textbf{Could we design an automatic method to generate visualizations from data-rich documents to support document-centric analysis?}

Because we aimed to support document-centric analysis, we first determined an appropriate style for the visualizations. We chose word-scale visualizations, a type of small, text-sized visualizations used to convey data insights in situ close to description~\cite{goffin2014exploring}. The small size of word-scale visualizations determined that they can seamlessly integrate into the existing format of data-rich documents, incurring minimal changes to a typical reading practice for data-rich documents. Moreover, prior works provided useful guidelines for the design, placement, and interaction of word-scale visualizations~\cite{goffin2015exploring, goffin2017Exploratory, goffin2020Interaction}, which we used to inform our design (Fig.~\ref{fig:teaser}).

In this study, we propose \textbf{\textit{GistVis}}, a framework for automatically generating word-scale visualizations from data-rich documents. We based our design on the formative findings regarding the narrative features of data-rich documents (Sec.~\ref{sec:formative-study}). Then, we designed GistVis as a proof-of-concept framework to which we applied several constraints to simplify the nuances in narrative features among data-rich documents. To provide flexibility for future expansion (Sec.~\ref{sec:gistvis-method}), we split the transformation process from text to visualization into four independent steps: Discoverer, Annotator, Extractor, and Visualizer (Fig.~\ref{fig:algorithm-pipeline}). For the first three processing stages, we utilized the capabilities of large language models (LLMs) through prompt chaining~\cite{wu2022ai} and injected visualization knowledge~\cite{wang2020datashot, amar2005lowlevel, chen2009effective} to steer LLMs to generate \textit{data fact specification}, an intermediate data representation to encode data insights in text documents. We applied a visualization design knowledge-driven approach in stage four to ensure predictability in the generated visualizations.

To assess the performance of GistVis, we conducted a comprehensive evaluation involving both a technical evaluation of the Discoverer and Annotator modules (Sec.~\ref{sec:technical-eval}) and a user study with 12 participants (Sec.~\ref{sec:eval-user-study}). Our technical evaluation revealed decent performance in both segmenting data insights and inferring the type of data insights. Meanwhile, users perceived the visualization generated by GistVis as useful—results showed a decrease in workload reading with GistVis compared to a Plain Text condition.

In summary, our main contributions are as follows:
\begin{itemize}
    \item We conducted a formative study where we collected a corpus of data-rich documents and investigated the narrative patterns of data-rich documents.
    \item We proposed GistVis\footnote{We document our implementation at \url{https://github.com/Motion115/GistVis}}, a framework driven by LLM and visualization knowledge for automatically generating word-scale visualizations from data-rich documents. %
    \item We demonstrated the utility of GistVis via a technical evaluation and a user study with 12 participants using real-world data-rich documents. Results indicated that GistVis could generate satisfactory and effective word-scale visualizations to support document-centric analysis of data-rich documents.
\end{itemize}

\section{Related Work}
We review three categories of previous works pertinent to our research question of designing an automatic method to generate word-scale visualizations for document-centric analysis. Specifically, we review 1) the definition, design, and application of word-scale visualizations, 2) related automatic visualization generation methods, and 3) how visualizations and texts could be integrated to enrich the reading experience.

\subsection{Word-scale Visualization}
Word-scale visualization has been discussed in the literature under several expressions, such as sparklines~\cite{tufte2006beautiful} and word-scale visualization~\cite{goffin2014exploring}.
\citet{tufte2006beautiful} introduced sparklines as concise, high-impact graphics that fit a text's typographic space. \citet{goffin2014exploring, goffin2017Exploratory} expanded Sparklines with word-sized visualizations, allowing more flexible integration of graphics and text.
Though terms may differ, one common theme between sparklines and word-scale visualizations is using text-sized graphics to augment existing documents. This study uses word-scale visualization to cover word-scale data visualization, text-sized typography, and text-sized graphics. 

In addition to discussing the word-scale visualization design space, researchers have also explored the potential of the application of word-scale visualizations~\cite{hoffswell2018Augmenting, perin2013soccerstories, brandes2013gestaltlines, beck2016visual}. For example, \citet{hoffswell2018Augmenting} implemented a design space that uses word-scale visualizations to augment the code reading experience. The SportLines interface~\cite{perin2013soccerstories} used word-scale visualizations to show the phases of players to support visual soccer analysis. These studies are formative to GistVis, as we employ the best practices integrating word-scale visualizations with text. Although prior work has explored many application scenarios, few have automated word-scale visualization generation for data-rich documents. This work contributes to the research of word-scale visualizations by proposing an automatic pipeline that generates word-scale visualizations to augment existing data-rich documents.

\vspace*{-2pt}
\subsection{Automatic Visualization Generation}
Research has explored various automatic visualization techniques, especially using tabular data as input~\cite{wang2020datashot, shi2021calliope, dibia2023lida}. For example, \citet{wang2020datashot} proposed the Datashot system that included a fact sheet generation pipeline based on a taxonomy of data facts~\cite{chen2009effective}. %
More recently, \citet{dibia2023lida} proposed the LIDA (Visua\underline{li}zation an\underline{d} Infogr\underline{a}phics) tool that incorporated large language models (LLMs) in the infographics generation process from tabular datasets.

A subcategory of automatic visualization revolves around generating visualizations based on textual contexts~\cite{wu2022ai4vis}. Under this category, we can broadly identify two ways of automatic visualization generation:
1) using external data sources and 2) using internal data sources (text description, tables, etc.). Using external data sources, NewsViews~\cite{gao2014newsviews} generated interactive visual maps through a table database. Contextifier~\cite{hullman2013contextifier} automatically created a stock timeline graph with annotations through sourcing external news corpus and APIs. Using internal data sources, Text-to-Viz~\cite{cui2020texttoviz} used regular expressions to detect data elements for generating proportional infographics. Charagraph~\cite{masson2023Charagraph} created in situ visualizations from data-rich paragraphs to support exploratory analysis of statistical data described in the text. However, those methods primarily used rule-based approaches to extract data insights from data-rich documents, and we need further work to capture semantically implied data insights in text descriptions.

GistVis categorizes as a context-based automatic visualization technique and uses data solely from internal sources. GistVis extends previous automatic visualization research by utilizing LLMs as a proxy to capture and extract data insights from text descriptions.
We employed a data fact-based approach~\cite{chen2009effective} inspired by the automatic methods mentioned above and engineered LLMs to use such visualization knowledge to generate word-scale visualizations.

\vspace*{-2pt}
\subsection{Visualization + Text}
\label{subsec:relatedwork-vistext}
Research has demonstrated the effectiveness of dynamically connecting text segments to visualizations to enhance the reading experience~\cite{latif2022kori, bromley2024dash, masson2023Charagraph}. Prior work has proposed crowdsourcing or mixed-initiative methods to create text-visualization connections. For example, \citet{kong2014extracting} used crowdsourcing to extract references between charts and text. For mixed-initiative methods, \citet{latif2022kori} proposed the Kori system, an interactive authoring tool synthesizing text and visualization. The DASH (\underline{D}ata \underline{A}nalysis using \underline{S}emantic \underline{H}ierarchies) system~\cite{bromley2024dash} leveraged LLMs to introduce semantic levels~\cite{lundgard2022accessible} into the bidirectional analysis between text and charts. The Charagraph system~\cite{masson2023Charagraph} used regular expressions to extract statistical data from a user-selected text domain to facilitate the understanding of statistical data in data-rich documents.

GistVis incorporates the practice of linking text with visualization by supporting two types of text-visualization interaction: 1) linking entities in text descriptions to visual elements in word-scale visualizations and 2) a hover tooltip that provides essential data insights of the selected word-scale visualization. Meanwhile, GistVis extends prior research by implementing text-visualization interaction on a word-scale setting. We expect the application of text-word scale visualization interaction could provide additional data context for users while reading. 

\section{Survey on Data-rich Documents}
\label{sec:formative-study}

To better motivate the design of GistVis, we conducted a formative study based on a corpus of data-rich documents to understand the narrative features of data-rich documents. In this section, we first describe how we collected the data-rich document corpus. Then, we describe how we derived the codes we used for our deductive coding process based on prior visualization literature. Lastly, we present the findings of our study and outline several constraints we took in the design of the GistVis automatic generation pipeline.

\subsection{Corpus Collection}
Data-rich documents are widely adopted in various domains; thus, many source documents are available online. Meanwhile, data-rich documents can also come in various genres, for instance, data journalism~\cite{stalph2018classifying} and scientific articles~\cite{beck2017wordsized}. Since it is more common for the general public to access data journalism, we included only data journalism in our corpus. Specifically, data journalism refers to news reports that contain rich data content (e.g., ~\cite{data-news-example, us-k12-education}).

Three researchers, who were briefed on the definition of data journalism and familiar with data visualization, applied a snowballing approach to collect data-rich news articles to construct the corpus. During the collection process, we attempted to cover a wide variety of topics while we labeled the topic for each data article referencing Stalph's topical classification codes~\cite{stalph2018classifying}. Ultimately, we collected 44 data-rich news articles that we labeled using six topical codes: \textit{Politics} (19.30\%), \textit{Society} (19.30\%), \textit{Business} (14.04\%), \textit{Culture} (15.79 \%), \textit{Local} (8.77\%) and \textit{Other} (22.81\%). All the articles were written in English.
We documented the source link of the documents in the supplementary material.

\subsection{Qualitative Analysis}
The first step in designing an automatic method for generating word-scale in situ visualizations is ascertaining the types of insights conveyed through text description. To obtain the candidate insight types, we coded the whole corpus deductively to 1) delimit the segment that contains data narrations and 2) code the segment with one or multiple insight types. Specifically, we coded the insight types referring to the facts taxonomy~\cite{chen2009effective}, which provided formal definitions that distinguished different types of data insights (e.g., value, distribution, difference, etc.). Recent research has successfully applied this taxonomy in the design of several automatic visualization techniques~\cite{wang2020datashot, shi2021calliope, chen2024chart2vec}.

Although the fact taxonomy is designed to reflect the attribute of the data insights independently across datasets or applications, we excluded several fact types less likely to be captured from text descriptions or conveyed through word-scale visualizations to simplify the coding process. First, limited information in text descriptions prevents us from obtaining complete tabular datasets. Therefore, we excluded data fact types commonly derived from data transformation on complete datasets (e.g., aggregation, anomaly). Secondly, we considered the possible display of the data fact type and excluded fact types that might be hard to read in a word-scale setting. For example, types such as association are commonly represented with chart types like Treemaps and Sankey Diagrams. %
Treemaps or Sankey Diagrams require much more space than what word-scale visualization could offer.

Ultimately, we had eight data fact types: comparison, trend, rank, proportion, extreme, value, distribution, and categorical. Through our coding process, we observed that \textbf{data-rich documents could cover many data fact types, but several data fact types are more common than others (O1)}. The frequency of occurrence and definition of the eight fact types are as follows:
\begin{itemize}
    \item \textbf{Value} (33.23\%; 332): The value type is one or multiple numerical values in a sentence retrieved under some specific criterion. For example, ``\textit{Sources report that almost 10 million migrants have crossed into the country.}''
    \item \textbf{Trend} (17.34\%; 168): The trend type is a general tendency of one data attribute over time. For example, ``\textit{In the last quarter of 2023, EV sales were up 40\% from the same quarter a year before.}''
    \item \textbf{Comparison} (16.72\%; 162): The comparison type measures the difference in value between two or more entities over a shared breakdown. For example, ``\textit{EVs create 3,932 pounds of carbon per year, compared to 11,435 for gas-powered vehicles.}''
    \item \textbf{Proportion} (15.07\%; 146): The proportion type expresses how much one or multiple attributes comprise the sum. For example, ``\textit{Around 60\% of Mexico is experiencing moderate to exceptional drought.}''
    \item \textbf{Extreme} (8.46\%; 82): The extreme type is the maximum or minimum of one specific attribute. For example, ``\textit{The highest mountain in the world is Mount Chumolongma at 8848 meters.}''  
    \item \textbf{Distribution} (3.61\%; 35): The distribution type demonstrates the numerical values shared over a specific breakdown. For example, ``\textit{Monocrystalline silicon production increased by 31.6 percent, photovoltaic cell production increased by 45.6 percent, NEV increased by 46.3, and wind power generator increased by 66.4 percent.}''
    \item \textbf{Rank} (3.61\%; 35): The rank type demonstrates an order or a sorted sequence over a specific breakdown. For example, ``\textit{This figure means the province contributed the second-highest GDP in China in 2023 only following South China's Guangdong Province with 13.57 trillion yuan.}''  
    \item \textbf{Categorical} (1.96\%; 19): The categorical type is data attributes with a certain joint feature. For example, ``\textit{The employment gains were concentrated in only three sectors: health care, leisure and hospitality, and government.}''
\end{itemize}

While identifying the data fact types, we also delimited the text segment that conveys the fact types. The criterion for extracting the text segments is to find the shortest possible sentence collection that coherently conveys a data insight. In the following, we refer to this shortest text segment that conveys one data insight as the \textbf{unit segment}. We found the majority of (84.21\%, 512) unit segments contain only one sentence, while the maximum count of sentences in a unit segment is 6. This suggested that \textbf{the majority of data insights occur within one sentence, but some cases exist where data insights span multiple sentences (O2).} Meanwhile, we did not consider cross-paragraph unit segments.

Note that because we did not further break sentences into fragments, each unit segment could, in theory, contain multiple data fact types. In alignment with the fact taxonomy, we refer to this scenario as compound fact. We found 46.05\% (280) unit segments in our corpus fall within the compound fact. The above data revealed that \textbf{many unit segments contain only one data insight, yet a good portion do contain multiple insights (O3)}, which could be further divided or interpreted in various ways. However, considering our goal in facilitating document-centric reading, we chose not to further break down the original sentence structure to avoid an overdose of visualizations. Hence, we decided to keep unit segments as our smallest analysis unit.

\subsection{Design Implications}
\label{subsec:formative-constriants-implications}
The above analysis revealed a rather complicated design space to fulfill, and designing a comprehensive automatic method that generates word-scale visualizations based only on text descriptions is challenging.
Therefore, as an initial attempt, we decided to apply several constraints driven by our observations from the corpus study to provide a better definition of our tasks.
We applied the following constraints (C1 - C3) when designing our automatic generation method:
\begin{enumerate}
    \item[\textbf{C1}] \textbf{We design our automatic generation method and visualization for more frequent data fact types (derived from O1).} Specifically, we support the data fact types of value, proportion, comparison, trend, rank, and extreme, the top six types in our corpus. We integrate distribution into the comparison category because they could share similar visual encoding. Additionally, we exclude categorical because of its low occurrence frequency.
    \item[\textbf{C2}] \textbf{We design our automatic generation method to be compatible with situations where a data insight can be extracted from the text in one unit segment (derived from O2).} This constraint ensures that we cover scenarios where data insights occur in consecutive sentences. Additionally, this constraint ensures that we can generate data-driven visualizations, which is the majority word-scale visualization type (79.5\%) created by human designers~\cite{goffin2017Exploratory}.
    \item[\textbf{C3}] \textbf{We design our automatic generation method to display only one data fact type (i.e., not concurrently process compound fact) (derived from O3).}  Although a moderate number of unit segments are compound facts, as a first step, we intend to evaluate the feasibility of generating one data fact type first before generalizing the method to support more complicated scenarios.
\end{enumerate}

On top of those constraints, we intended to leave the implementation flexible, which grants extensibility and enables further expansion and optimization. Considering all observations (O1 - O3), constraints (C1 - C3), and the extensibility objective, we bear three overarching design goals (DG1 - DG3):

\begin{enumerate}
    \item[\textbf{DG1}] \textbf{Establish a uniform data structure to encode both plain text and data insights.} The architecture should treat the data insights at the same level as plain text, ensuring that the word-scale visualization is generated from one unit segment from a document-centric perspective.%
    \item[\textbf{DG2}] \textbf{Apply modular design principle.} Provide abstraction to the automatic word-scale visualization generation process to support plug-and-play property for each processing module to optimize performance and extensibility.
    \item[\textbf{DG3}] \textbf{Design reusable and expressive word-scale visualization components that support interactive document-centric analysis.} The visualization design should be succinct and clear to express insights into various data fact types. They should also adhere to the space limitations of the word-scale setup while offering interaction between text and visualization to improve its clarity. Considering extensibility, the word-scale visualization components should be independent of data insights to support the easy integration of additional word-scale visualization designs.
\end{enumerate}

\section{GistVis}
\label{sec:gistvis-method}
The following sections describe the design of GistVis. We first define the uniform data representation schema \textit{data fact} (\textbf{DG1}, Sec.~\ref{sec:gistvis-gistfact}). Then, we provide a detailed description of the purpose of each module in the GistVis computational pipeline, how it could support plug-and-play property (\textbf{DG2}, Sec.~\ref{sec:computational-pipeline}), and the design choices of our word-scale visualizations (\textbf{DG3}). Lastly, we introduce the implementation details of the current iteration of GistVis (Sec.~\ref{sec:implementation-detail}).

\subsection{Data Fact}
\label{sec:gistvis-gistfact}

We define \textit{data fact} as the uniform data structure to encode all text content, either with or without data insights. Under this definition, we characterize \textit{data fact} as a declarative intermediate data structure that encapsulates all the key information required to generate word-scale visualizations. We formulate \textit{data fact} as a 2-tuple:
\begin{equation}
\mathTextItalics{fact}\, \coloneq \left \{ \mathTextItalics{unitSegmentSpec, dataSpec?} \right \} \nonumber
\end{equation}
where \textit{unitSegmentSpec} records key information related to the \textbf{unit segment}. Meanwhile, \textit{dataSpec} is a list that contains the data restored from the information provided by the unit segment. Because not all unit segments contain data insights, \textit{dataSpec} is an optional attribute in the tuple and could be used to distinguish between plain text segments and data insight segments. For clarity, we use the \textit{Typescript} notation (question mark) to represent optional elements. In the following, we explain the design of \textbf{unit segment specification} (\textit{unitSegmentSpec}) and \textbf{data specification} (\textit{dataSpec}).

\subsubsection{Unit Segment Specification}
\label{sec:gistvis-unitSegmentSpec}
We define unit segment specification as a 4-tuple:
\begin{equation}
    \mathTextItalics{unitSegmentSpec}\, \coloneq \left \{
        \mathTextItalics{type, context, attribute?, position?}
    \right \} \nonumber
\end{equation}

\paragraph{Type}
Type contains seven candidate types, including the six data insight types (i.e., value, proportion, comparison, trend, rank, and extreme) we selected from the facts taxonomy (see Sec.~\ref{sec:formative-study}), and one ``no type'' to represent plain text. 

\paragraph{Context}
Content is where we store the original text snippets that are unit segments. In this work, we define a unit segment as one or multiple sentences that collectively convey one data insight or are similar in semantics. We implement this definition to support situations where data insights span multiple adjacent sentences.

However, it is worth noting that we defined unit segments based on a strong assumption that relevant information of one data insight is described in sequential order, proximate in position, and contained within the same paragraph. Cases might exist where the same data insight could occur in numerous places across the document~\cite{goffin2020Interaction}. We argue that word-scale visualizations designed for document-centric analysis might not be the optimal solution to convey such information. Thus, we only focus on performing paragraph-level segmentation.

\smallskip
The above two entries (\textit{Type}, \textit{Context}) record contextual information directly related to the text and are ubiquitous for all unit segments. Meanwhile, \textit{Attribute} and \textit{Position} are data-related auxiliary information of the text and relevant to specific \textit{Types}. Referencing the four-level model of the semantic content of visualization proposed by \citet{lundgard2022accessible}, we viewed \textit{Attribute} and \textit{Position} to represent L2 - L3 (statistical concepts and relations, perceptual and cognitive phenomena) information and L1 (elemental and encoded properties) information respectively. Because \textit{Attribute} and \textit{Position} convey key semantic insight about the data and are closely related to text descriptions, we leave those attributes in the unit segment specification.

\paragraph{Attribute}
Attribute is an optional entry explicitly designed for data types extreme and trend. The candidate options for attributes include ``increasing'' and ``decreasing'' for the trend type and ``maximum'' and ``minimum'' for the extreme type. The attribute information is a supplementary constraint in the visualization generation process to correct potential errors in the generated visualizations and reflect the semantics of the text description.

\paragraph{Position}
The position is an optional entry to handle the extreme data fact type. It represents the original text description that should be highlighted to provide contextual information about the maximum or minimum values. For example, if we want to augment "\textit{the maximum of sales for company A,}" it is more informative to label the entire phrase rather than just "company A" (the standard highlight practice for other data fact types). If position is available in the extreme data fact type, the position attribute would override the default practice and ensure the entire phrase is highlighted.

\subsubsection{Data Specification}
\label{sec:gistvis-dataspec}
\begin{figure}[tb]
  \centering
  \includegraphics[width=0.8\linewidth]{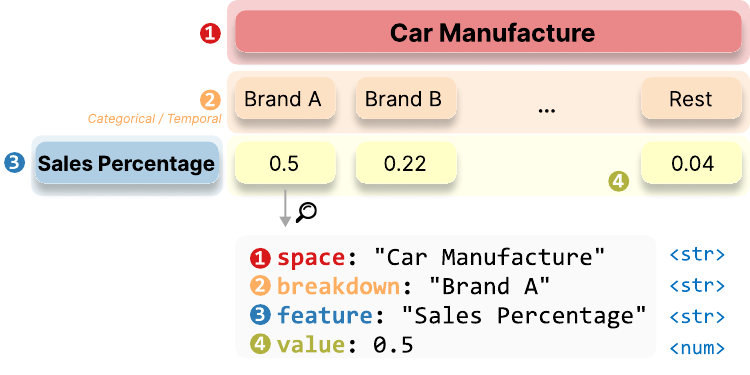}
  \caption{Each element in data specification consists a four-tuple, space \ding{182}, breakdown \ding{183}, feature \ding{184} and value \ding{185}.}
  \label{fig:dataspec-tableview}
\end{figure}

While Unit Segment Specification characterizes the textual content and higher-level data insights, data specification represents the raw data elements reconstructed from the textual content. Data Specification is designed to be an analogy of tabular datasets (Fig.~\ref{fig:dataspec-tableview}), which we define in a four-tuple:
\begin{equation}
    \mathTextItalics{dataSpec}\, \coloneq \left \{ 
        \mathTextItalics{space, breakdown, feature, value}
    \right \}[] \nonumber
\end{equation}

Specifically, \textbf{space} is a facet of analysis with a given text description. For example, if a sentence describes the market share of different car manufacturers, the analysis space would be ``car manufacture'' (Fig.~\ref{fig:dataspec-tableview} \ding{182}). Meanwhile, \textbf{breakdown} is a set of temporal or categorical data fields in which data are further divided under the space. For example, the brand name, like ``Brand A'' (Fig.~\ref{fig:dataspec-tableview} \ding{183}), would be the breakdown for ``car manufacture''. \textbf{Feature} is the measurement of breakdown. For example, we could measure the sales percentage for each manufacturer (Fig.~\ref{fig:dataspec-tableview} \ding{184}), a feature derived from annual sales of car manufacturers. Lastly, \textbf{value} is a numerical data field that could be retrieved from a combination of breakdown and feature. For example, the ``sales percentage'' of ``Brand A'' is 0.5 (Fig.~\ref{fig:dataspec-tableview} \ding{185}). All data attributes are required for each data specification entry, with the only exception being the ``not a number'' (NaN) value attributes. Cases exist when the unit segment describes a semantic data insight (e.g., increasing or decreasing for the trend type), and we make ``not a number cases'' a special condition for GistVis to process.

In the following section, we describe how the fields in the data fact specification are filled using the GistVis pipeline.

\subsection{Computational Pipeline}
\label{sec:computational-pipeline}
\begin{figure*}
    \includegraphics[width=1\textwidth]{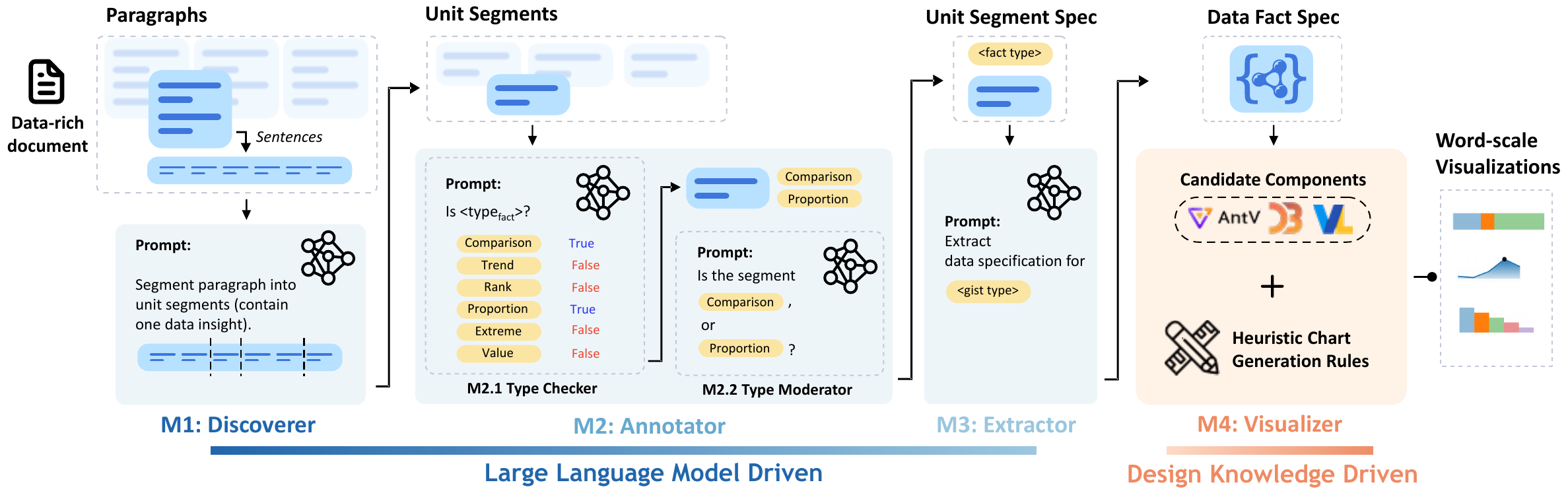}
    \caption{The GistVis pipeline consists of four modules: Discoverer (\textbf{M1}), Annotator (\textbf{M2}), Extractor (\textbf{M3}), and Visualizer (\textbf{M4}). Data flows through the four modules sequentially, where a large language model captures the insight of the data-rich document (\textbf{M1-M3}). Visualizer (\textbf{M4}) maps the captured insight into interactive visualizations, populated in situ in the text document at word scale.}
    \label{fig:algorithm-pipeline}
\end{figure*}

Based on the data requirement of data fact, we proposed the GistVis pipeline to automatically transform data-rich text descriptions to word-scale visualizations. We utilized both LLM-based and design knowledge-driven approaches to achieve word-scale visualization generation. Specifically, we decomposed the generation process into four stages: gist discovery, fact type annotation, fact specification extraction, and fact visualization. We capitalized on LLMs' natural language understanding capability for the first three stages and applied the prompt chaining strategy~\cite{wu2022ai} to transform natural language descriptions into data facts. Meanwhile, the fact visualization stage uses a simple heuristic-driven approach to map data facts to interactive visualization components. As an automated process, all prompts and visualization heuristic rules are readily coded into the pipeline (see Supplementary Material for detailed prompt design). Fig.~\ref{fig:algorithm-pipeline} shows the GistVis computational pipeline, which consists of four modules respective to the four stages above: Discoverer (\textbf{M1}), Annotator (\textbf{M2}), Extractor (\textbf{M3}) and Visualizer \textbf{(M4)}.

\subsubsection{Discoverer}
The first step in the GistVis pipeline is to divide paragraphs into unit segments. Discoverer (Fig.~\ref{fig:algorithm-pipeline}.~\textbf{M1}) leverages the zero-shot capability of LLMs~\cite{brown2020language} to perform the segmentation process. We restrain LLMs from identifying the shortest unit segment possible to better pair the text descriptions with in situ word scale visualizations. The prompt contains the six fact types to provide LLMs with more detailed segmentation requirements. Meanwhile, although LLMs are informed of the data fact types in the instructions, we do not label the data fact type at this stage to keep the task simple.
Moreover, the segmentation process shortens the context length, filtering out excess information for the subsequent data fact type labeling step. Additionally, to reduce hallucinations from LLMs during the segmentation process, we instruct the LLM to keep the text description ``as is'' without modifying the original text description or punctuation.

\subsubsection{Annotator}
The Annotator module (Fig.~\ref{fig:algorithm-pipeline}.~\textbf{M2}) aims to fill in the \textit{Type} field in the data fact specification for each segment from the prior module. Since LLMs are well calibrated to answer multiple choice and true/false questions~\cite{kadavath2022language}, we formulate the data fact type annotation as a two-stage question-answering (QA) problem. In the first stage, we ask LLMs to make a true/false judgment on whether the given segment belongs to a specific data fact type (\textbf{Type Checker}). Then, since one segment can be classified into multiple data fact types, we perform another round of prompting (\textbf{Type Moderator}) in a multiple-choice format to determine the most appropriate data fact type for each segment.

\paragraph{Type checker}
Type Checker identifies all possible fact types for each segment. To incorporate visualization knowledge into LLMs, we applied the few-shot in-context learning prompting paradigm~\cite{brown2020language}. We constructed seven individual reasoners for each of the seven fact types. Specifically, following a task description informing LLMs to return whether or not the segment can be classified as a given fact type, we define the fact type followed by three examples (two positive, one negative), allowing the LLMs to capture the definition of data fact type and forming a 3-shot prompt. After running through all seven reasoners, we record all the possible data fact types in a list for further moderation. If all reasoners responded false, the segment would be labeled a text-only segment (i.e., no type) and not proceed for further analysis.

\paragraph{Type moderator}
Taking the output from the Type Checker, the Type Moderator determines the most suitable data fact type for a given segment. We formulate the prompt in a multiple-choice pattern, with the options drawn from the previous type-checking step. Due to the varying number of possible candidates, we do not include examples and only reiterate the definition for the candidates. After the moderation step, we narrow the data fact label to only one label and fill the type field in the data fact specification in preparation for the data extraction phase.

We annotated data fact types in two instead of one round for the following reasons. First, we limit the input length and avoid excessive length prompts that could potentially harm inference capability~\cite{li2024longcontext}. Splitting the annotation task into two rounds would save space—we avoided all the samples for each data type occurring in the same prompt. Secondly, we consider the extensibility of the GistVis pipeline. Splitting the process into two stages would ensure that adding custom data fact types requires no more than writing additional prompts for the new data fact type. We also justified this design decision with an ablation study in our technical evaluation (Sec.~\ref{subsec:quant-eval-annotator}).

\subsubsection{Extractor}
Since each data fact type has different requirements, the Extractor module (Fig.~\ref{fig:algorithm-pipeline}.~\textbf{M3}) applies a case-by-case extraction strategy based on the data fact type generated from the Annotator module when extracting the data specifications. We formulate the prompts based on the visualization specification requirement described in Sec.~\ref{sec:gistvis-dataspec}.

Compared with the methods that applied regular expression to identify numbers in a text description, LLMs allow more flexibility in how the data is presented. For example, when numbers are not expressed numerically (e.g., 10 thousand, ten thousand), regular expressions would fail in extracting the correct underlying data. Instead, we prompt LLMs to convert non-numerical expressions of numbers to their numerical form (i.e., 10000), thus expanding data extraction capability. We also use a number parser to further transform all extracted data into a numerical form.

\subsubsection{Visualizer}

\begin{figure*}[tb]
  \centering
  \includegraphics[width=\linewidth]{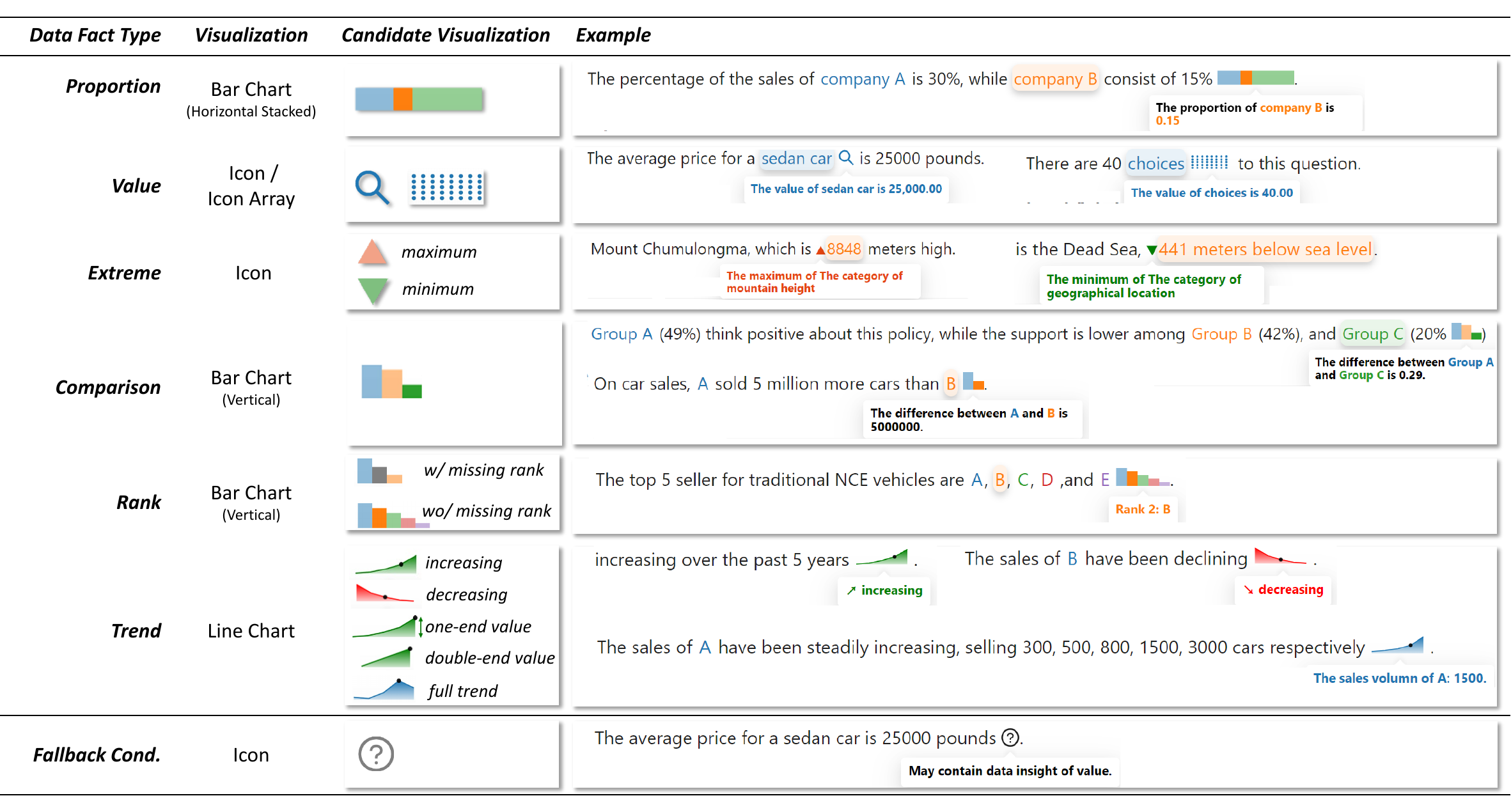}
  \caption{A collection of 14 candidate visualizations and the corresponding chart type for each data fact type. The \textbf{Example} column shows the effect of the appearance of word-scale visualization in data-rich documents. We present the examples when the mouse hovers over the word-scale visualization of focus.}
  \label{fig:visualization-design}
\end{figure*}

Visualizer (Fig.~\ref{fig:algorithm-pipeline}.~\textbf{M4}) applies a chart-based approach (as opposed to the grammar of graphics~\cite{wilkinson2012grammar} approach) in visualizing data facts. We took the chart-based approach based on the observation that various business visualization toolkits~\cite{g2plot-2024, echarts-2024} use this paradigm. 

\paragraph{Visualization Design}
We implement three basic chart types (bar chart, line chart, and icon array) and use their variants with relevant icons to tailor to the needs of different data fact types. The variants of the three chart types and icons constructed a design space including 14 candidate visualizations to represent different conditions of data insights (Fig.~\ref{fig:visualization-design}). Notably, we use horizontal stacked Bar Charts to represent proportion. Although prior works have applied Pie Charts (e.g.,~\cite{goffin2017Exploratory, huth2024eye}), we argue that the limited height could make discerning the angle difference between categories difficult. Thus, in this work, we attempt to capitalize on the relatively ample space on the horizontal dimension and use length, a more effective visual channel than angle~\cite{munzner2014visualization}, to encode proportional data. We also made several hard constraints to facilitate the readability of word-scale visualizations. For example, we limit the maximum rank for visualization to 10, avoiding situations of ultra-long vertical bar charts if the value of the ranking is large. We made this decision based on the fact that only one in 35 rank insights we identified in our corpus (Sec.~\ref{sec:formative-study}) included a rank above 10. Moreover, the definition of the rank fact type implied a sorting process on the dataset such that the rank is relative, leading to the inclination of reporting ranking with single-digit numbers. Meanwhile, to avoid situations such as rank overflow, we also designed a fallback condition, presenting a question mark icon to indicate the unit segments that might contain data insights not properly presented by GistVis. We expect that users could better perform analytical activities~\cite{amar2005lowlevel} with data-rich documents through the locally aggregated views of data in the form of word-scale visualizations.

\begin{table*}[tbp]
\small
  \centering
  \caption{The visualization knowledge the Visualizer (M4) module assumes when parsing data fact specification. C and T represent categorical and temporal data types, whereas N represents the numerical data type. The fallback condition will be hit if data is missing in the data fact specification.}
    \begin{tabularx}{\linewidth}{cccccX}
    \toprule
    \textbf{Fact Type} ($t$) & \textbf{Breakdown} ($b$) & \textbf{Feature} ($f$) & \textbf{Attribute} ($a$) & \textbf{Position} ($p$) & \textbf{Tooltip Syntax} \\
    \midrule
    Proportion & C/T   & N     & \ding{53}     & \ding{53}     & The proportion of $\{b_i\}$ is $\{v_i\}$. \\
    Value & C/T   & N     & \ding{53}     & $\geq$1 & The value of $\{b_i\}$ is $\{v_i\}$. \\
    Extreme & C/T   & N     & \makecell[c]{maximum/\\minimum} & $=$1  & The $\{a_i\}$ of $\{b_i\}$. \\
    Comparison & C/T   & N     & \ding{53}     & \ding{53}     & The difference between $\{b_i\}$ and $\{b_j\}$ is $\{| v_i - v_j|\}$. \\
    Rank  & C     & N     & \ding{53}     & \ding{53}     & Rank $\{v_i\}$: $\{b_i\}$ \\
    Trend & T     & N     & \makecell[c]{positive/\\negative} & \ding{53}     & \makecell[l]{$\{a_i\}$ \\ $\{f\}$ of $\{b_i\}$ is $\{v_i\}$. \\ The $\{a_i\}$ is $\{| v_i - v_j|\}$} \\
    \midrule
    (Fallback cond.)  & \ding{53}     & \ding{53}     & \ding{53}     & \ding{53}     & May contain data insight of $\{t\}$. \\
    \bottomrule
    \end{tabularx}
  \label{tab:gistvis-implementation}
\end{table*}

\paragraph{Linking Word-scale Visualizations with Text}
We also added interactive features to word-scale visualizations. The motivation behind making word-scale visualizations interactive is twofold: 1) to pack more information in word-scale visualizations, and 2) to enhance the reading experience by coupling visualization with text (Sec.~\ref{subsec:relatedwork-vistext}). 

Firstly, to pack more information within word-scale visualizations, we designed a drill-down interaction that pops up a tooltip every time users hover over the visualization. The tooltip contains basic data descriptions about the word-scale visualization, allowing users to grasp key information even when viewing the visualization standalone. We define a set of default syntax to describe the data insights for each data fact type according to the value types in the data fact specification (Table.~\ref{tab:gistvis-implementation}). Crucially, users access the tooltips on demand, minimizing their impact in obstructing a normal document reading process.

Secondly, to enhance the reading experience, we designed bidirectional interactions between text and word-scale visualizations. Specifically,
we highlight the ``entities'' of a sentence via matching document text with the set of extracted breakdowns in the data fact specification. When a specific visual element is selected, the corresponding entity synchronously lights up to show the correspondence between the entity and its value. Conversely, when a specific entity is selected, the corresponding visual element in the word-scale visualization would also light up to help users retrieve the entity related to the value (see Fig.~\ref{fig:visualization-design} Example column).

\subsection{Implementation}
\label{sec:implementation-detail}
We implement the GistVis pipeline using a typical web stack. For rendering the data-rich document augmented by GistVis, we use open-source libraries, including React\footnote{\url{https://react.dev/}} as the UI framework and D3.js~\cite{bostock2011datadriven} for rendering the word-scale visualizations in svg format. We took inspiration from prior jQuery-based word-scale visualization packages such as Sparklificator~\cite{sparklificator-package} and Piety~\cite{piety-package} and expanded their functionality using the React framework.

As for LLMs, we chose \texttt{DeepSeek-V2.5}~\cite{deepseek-ai2024deepseekv2}, an open-source\footnote{\url{https://huggingface.co/deepseek-ai/DeepSeek-V2.5}} Mixture-of-Experts (MoE) language model released by \texttt{DeepSeek}\footnote{\url{https://www.deepseek.com/en}}. We made this decision based on its decent performance at a low cost (overall cost less than 0.28\$ per 1M tokens). However, since we did not have the computation device to run such a large model, we resorted to the commercial API release of the exact DeepSeek model in our implementation. 
We then chained the processing steps with LangChain.js v0.1~\cite{langchainjs}, the JavaScript implementation of a popular framework for developing LLM-powered applications.

\section{Technical Evaluation}
\label{sec:technical-eval}
The effectiveness of GistVis depends on whether our computational pipeline can successfully extract key information from data-rich documents to create word-scale visualizations. Therefore, we conduct a technical evaluation to assess the performance of the GistVis submodules. Specifically, we provide quantitative results on the performance of the Discoverer and Annotator. Because Extractor and Visualizer is essentially a restoration process bound with information loss, we argue it would be challenging to provide reliable ground truth to evaluate the two modules. Hence, we demonstrate the performance of the Extractor and Visualizer with the user study results in the next section (Sec.~\ref{sec:eval-user-study}).

\subsection{Discoverer Evaluation}
The purpose of Discoverer is to delimit paragraphs into unit segments. In this section, we demonstrate the effectiveness of the Discoverer by comparing it with alternative methods that share the same algorithmic goal as the Discoverer.

\subsubsection{Experiment Setting}
We use the annotated corpus described in Sec.~\ref{sec:formative-study} to evaluate the performance of the Discoverer. Following the requirement of our computational pipeline, we annotated unit segments within the hard boundaries of paragraphs. We selected paragraphs that contained at least one data insight because we did not label unit segments for paragraphs without data insights. After filtering, our evaluation corpus included 640 paragraphs, each containing one or multiple unit segments. 

To provide a reference for comparison, we implemented three alternative methods. We first split paragraphs into sentences using the sentence tokenizer from the natural language processing toolkit nltk~\cite{bird2009natural} for all three alternative methods. Then, we applied different strategies to form unit segments. The first strategy, namely ``regex'', used regular expression to detect numbers inside a sentence. We identified a sentence as a unit segment if it includes a number. The second and third strategies involve using pre-trained language models, specifically BERT~\cite{devlin-etal-2019-bert}\footnote{\url{https://huggingface.co/google-bert/bert-base-uncased}} and sentence BERT~\cite{reimers-gurevych-2019-sentence}\footnote{\url{https://huggingface.co/sentence-transformers/all-MiniLM-L6-v2}}, to group similar sentences into unit segments. We first computed sentence embeddings, then computed the similarity between sentences, forming a unit segment if the similarity was above a certain threshold. We ran a grid search over multiple thresholds and reported the result of the best-performing threshold for each approach.

We benchmarked the performance of our Discoverer module against three alternative methods: regex, BERT, and sentence BERT, using the evaluation corpus mentioned above. %
We report the accuracy of segmentation for each condition. We define an accurate segmentation as making the same segmentation as our annotation. Our results would be a conservative indicator of the model's performance because we classify both over-segmentation and under-segmentation as incorrect, accepting only perfect matches. Over or under-segmentation of sentences without data insights would not impact the visualization result. Yet, we still require our algorithms to correctly delimit the boundaries to facilitate clear communication of our evaluation and compare different approaches.

\subsubsection{Result}
\begin{table}[tbp]
    \centering
    \caption{Comparison of segmentation accuracy for four candidate approaches.}
    \begin{tabular}{rrr}
    \toprule
       \textbf{Strategy}  & \textbf{Accuracy} & \textbf{Threshold} \\
    \midrule
       Regex  & 0.545 & - \\ 
       BERT &  0.611 & 0.81\\ 
       Sentence BERT & 0.600 & 0.68 \\ 
       Discoverer (LLM)  & \textbf{0.686} & - \\ 
    \bottomrule
    \end{tabular}
    \label{tab:discoverer-eval}
\end{table}

Results revealed moderate performance for all approaches, with the strategy applied in the Discoverer module performing the best (Table.~\ref{tab:discoverer-eval}). Specifically, the LLM-driven method we applied in the Discoverer module performed best, perfectly segmenting 68.6\% of the paragraphs. Meanwhile, the regex strategy fared worst, perfectly segmenting only 54.5\% of the paragraphs. The pre-trained language models' strategies gave a modest performance (61.1\% for BERT and 60.0\% for Sentence BERT). However, a trade-off of using the LLM-driven Discoverer is the processing time. Though the exact generation time through calling LLM APIs would be affected by internet speed and server load, we provide a rough expectation of this method's processing speed. We recorded an average of 5.22 seconds (SD=2.13s) to split each paragraph, with an average of 0.12 seconds (SD=0.03s) per-word processing time.

\begin{figure*}[tbp]
    \centering
    \includegraphics[width=0.85\linewidth]{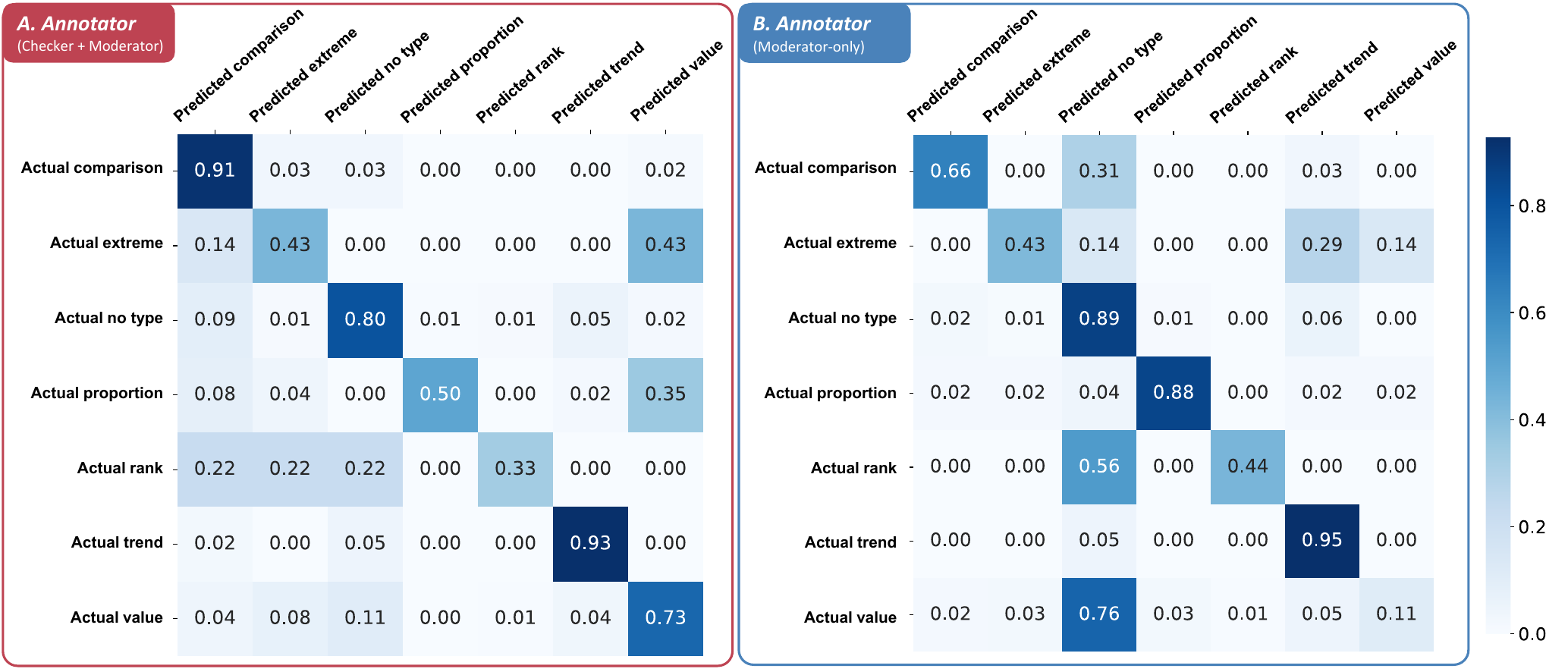}
    \caption{Normalized confusion matrices for data fact type annotation results. The left matrix (A) shows the result of our two-step Annotator (Type Checker + Type Moderator), while the right matrix (B) shows the result of the ablated condition (Type Moderator only). The horizontal axis denotes the predicted type, while the vertical axis indicates the actual type. The numbers on the diagonal line of this matrix represent the precision of classification for each category.}
    \label{fig:type-confusion-mat}
\end{figure*}

\subsection{Annotator Evaluation}
\label{subsec:quant-eval-annotator}

The goal of the Annotator is to label each unit segment with the corresponding data fact type. Thus, we demonstrate the Annotator's effectiveness by reporting its classification performance. We also justify the two-step Annotator design (Type Checker + Type Moderator) through an ablation study over a one-step design (Type Moderator only).

\subsubsection{Experiment Setting}
We use the annotated corpus described in Sec.~\ref{sec:formative-study} to evaluate the performance of the Annotator. To understand the performance of this single module, we assume the segmentation is correct and directly use the labeled unit segments as input. To suit the scope of the GistVis pipeline, we excluded types that are not yet supported and excluded meta-fact entries. After the above exclusion, we are left with 2676 unit segments as input. The dataset is highly imbalanced in its label because data insights are generally scarce even in data-rich documents. Specifically, there are 2355 unit segments without data fact (no type), 158 values, 41 trends, 58 comparisons, 7 extremes, 48 proportions, and 9 ranks, respectively. Because of the label imbalance, we report accuracy, and the weighted precision, recall, and F1-score to reflect the pipeline performance objectively. We also report the average annotation time per unit segment to provide a full picture of the performance of this module.

For our ablation study, we disabled the Type Checker in our Annotator module. Because the two-step Annotator does not pass ``no type'' segments to the Type Moderator, we slightly modified our prompt to support the output of the ``no type'' label to enable this comparison. The rest of the experiment was kept the same to eliminate confounding variables.

\subsubsection{Result}

Results revealed decent performance of the Annotator, achieving an overall accuracy of 0.79, a precision of 0.92, a recall of 0.79, and an F1 score of 0.84. Fig.~\ref{fig:type-confusion-mat}. A is the confusion matrix of the classification results. We identified several data fact types prone to misclassification, including rank, proportion, and extreme. For instance, the Annotator frequently misidentified proportion and extreme types as the value type. Since one or multiple values typically occur in proportion and extreme types, such misclassification exerts a minor impact during the visualization stage. However, visualizing proportion and extreme as the value type could potentially lead users to misunderstand the data insights. Additionally, rank was often misclassified as difference, extreme, or no type, which could undermine the expression of the rank insight during the visualization stage. We recorded an average inference time for each unit segment of 2.34 seconds (SD=2.46s). The large standard deviation in time was due to some inferences only passing one sub-stage (e.g., no type or only identified one data fact type). In contrast, others needed two passes to finalize the data fact type inference, which would significantly increase inference time.

At first glance, the ablated one-step condition (Type Moderator only) seemed to reveal comparable overall performance to the two-step Annotator, achieving an overall accuracy of 0.84, a precision of 0.89, a recall of 0.84, and an F1 score of 0.84. However, a breakdown in the classification performance of each category (Fig.~\ref{fig:type-confusion-mat}. B) revealed a significant pitfall of this design: the one-step condition was inclined to assign ``no type'' to unit segments. Specifically, the one-step condition significantly reduced the precision of the ``value'' type to only 0.11, an unacceptable result because the ``value'' type consists of as much as 33\% of the data insights in the corpus we collected. We attribute this difference between the two and one-step Annotator to the ``sifting'' effect of the Type Checker. The multiple parallel Type Checkers before the Type Moderator module filtered out most of the ``no type'' conditions and other unlikely data fact types, reducing the subsequent classification with fewer options. A smaller decision space would typically make classification easier. Another benefit of using a two-step design is that each Type Checker could be bespoke to achieve better performance. For example, we could design specific prompts using advanced prompting techniques such as In-context Learning~\cite{yao2024more} to bolster the performance further or even use different classification models for each data fact type.

\section{User Study}
\label{sec:eval-user-study}
To evaluate GistVis, we first demonstrated three target use cases related to its application. Then, based on low-level tasks~\cite{amar2005lowlevel} related to the use cases, we conducted a user study to understand the efficacy of GistVis and gather user feedback into improving GistVis.

\subsection{Use Cases}
\paragraph{Reading News Articles.}
Emily is keen to understand the current state of the healthcare industry, particularly the impact of a recent medical strike. To quickly capture the gist, Emily resorts to GistVis-generated visualizations to search for key data insights conveyed in the narration of this news article. She quickly captured the government's plan \includegraphics[height=10px]{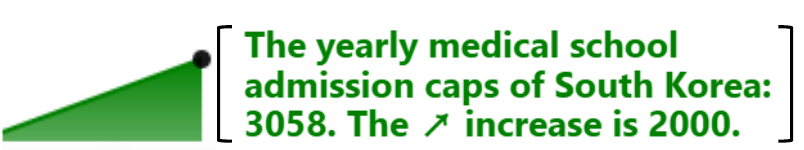} and the corresponding gap compared to other countries \includegraphics[height=10px]{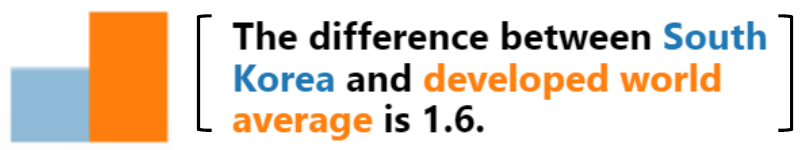} with the word-scale line charts and bar charts. She also discovered that junior doctors only account for roughly 35\% of total doctors through a word-scale stacked bar chart \includegraphics[height=10px]{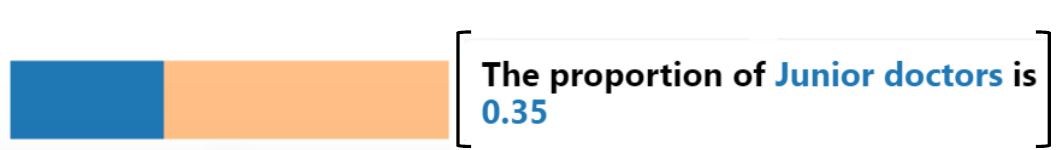}.

\paragraph{Reading Business Reports.}
As an investor, David wants to find key indicators of an automotive firm from a seasonal report. Using GistVis, he quickly found that the average price of electric vehicles had fallen by 0.9\% \includegraphics[height=10px]{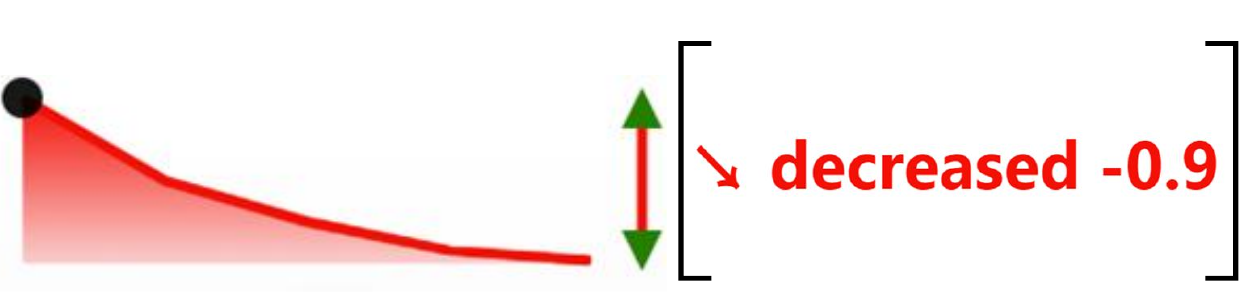}. He also captured the changes in vehicle supply compared to last year \includegraphics[height=10px]{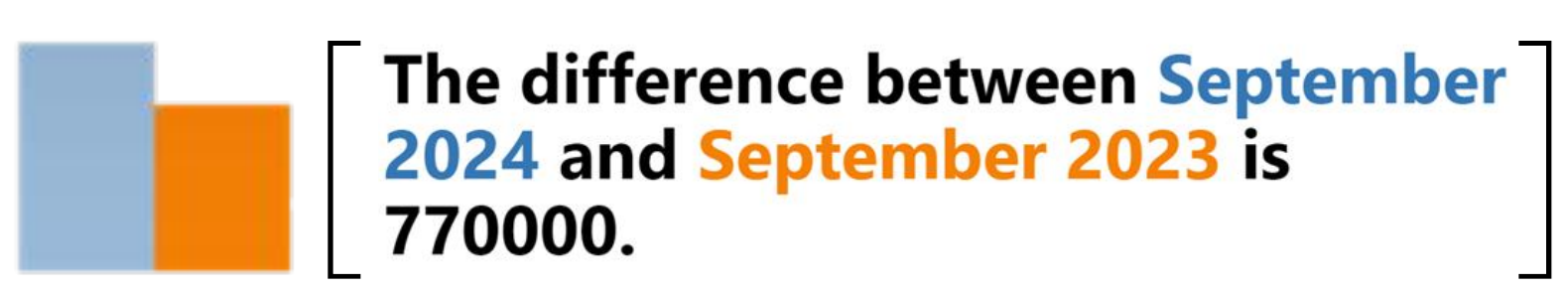} and the average value of EV incentives\includegraphics[height=10px]{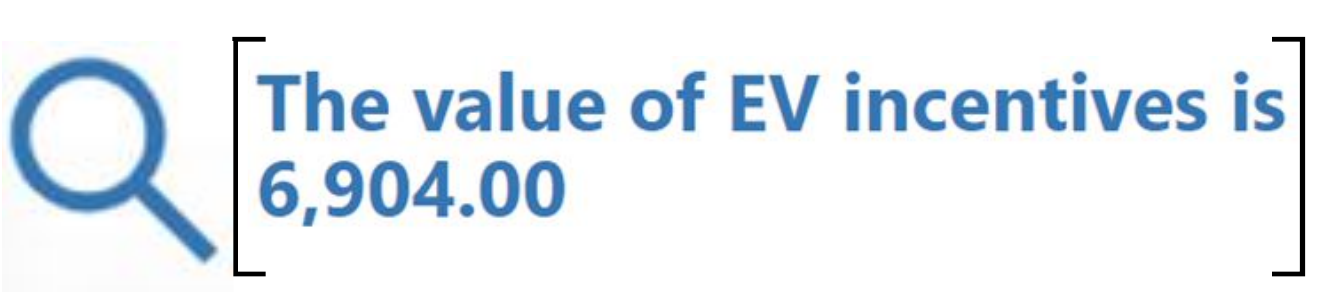} with the word-scale bar charts and icons. GistVis's visualizations highlighted the placement of data insights, which propelled him to navigate through all the key indicators swiftly.

\paragraph{Reading Academic Papers}
Sarah is a student passionate about exploring interdisciplinary topics. When reading a visualization paper, she quickly understood the key findings via GistVis: many participants used the generated charts \includegraphics[height=10px]{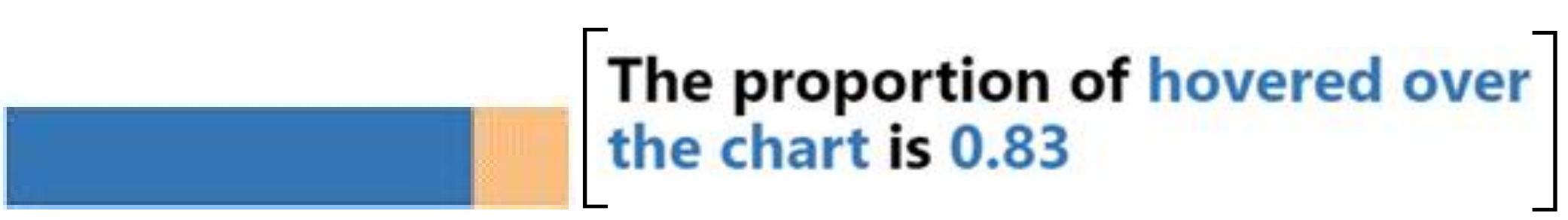}, and the new design elicited better user performance than the baseline condition \includegraphics[height=10px]{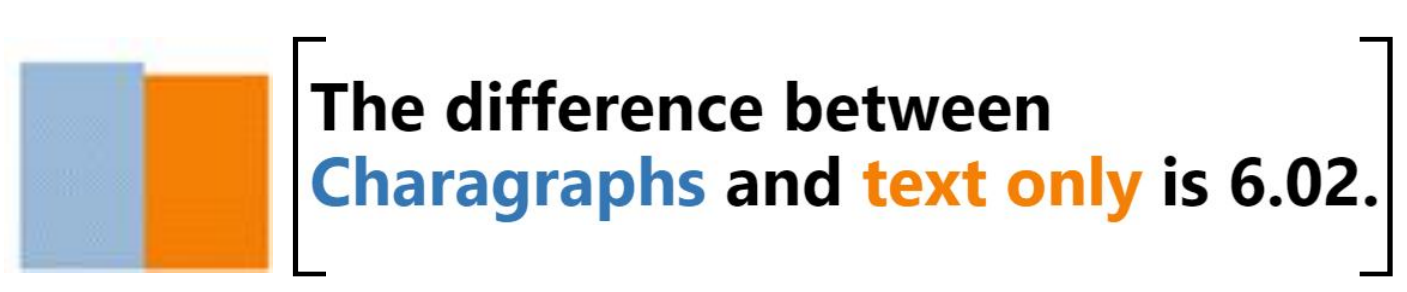}. By visualizing key data expressions in the paper, GistVis enables her to explore new topics with confidence.

\subsection{Goals, Conditions, and Metrics}
Based on the prior description of GistVis's use case, we aim to understand the following two facets of GistVis:
\begin{itemize}
    \item \textbf{Utility}: Can users better understand data-rich documents with GistVis-generated visualizations?
    \item \textbf{User Perception}: How do users perceive the new content generated by GistVis, including the visualization design, interaction model, generation quality, and its impact on the reading workload?
\end{itemize}

To understand whether GistVis provided incremental benefits to users, we compare reading with GistVis (condition \textit{GistVis}) to the traditional experience of reading data-rich documents with only the plain text (condition \textit{Plain Text}). We are aware that some data-rich documents have visualizations accompanying the textual descriptions. However, since we aim to understand whether GistVis benefits users, we decided not to include the original visualizations in those documents as they are neither word-scale nor widely applicable to all data-rich documents.

We devised a reading task and used several standardized tests to capture the utility and user perception of GistVis. For utility, participants were invited to respond to four multiple-choice questions related to the data presented in the text articles and one summary question. The four multiple-choice questions test users' ability to perform data analysis tasks~\cite{amar2005lowlevel} with data-rich documents under the two conditions.
For example, we phrase `Find Proportion' as `\textit{What is the percentage of [entity]?}' and the `Compare' task as `\textit{Which is the highest/second-highest among [entities]?}' We calculated the accuracy of the participant's response and compared the accuracy between each condition (see Supplemental Material for materials, questions, and answers). We also collected interaction logs containing participants' frequency of visiting the word-scale visualizations and the dwell time in the GistVis Condition. We interpret this data by comparing the interaction log with the position of the word-scale visualization in which the answer lies. We seek to understand if the generated visualizations helped or interfered with participants' answers. In addition, after completing the four multiple-choice questions, participants were required to summarize the document's central argument. The results of the summary questions were not included in our evaluation but served as a stimulus to ensure participants finished reading the whole article. We recorded the participants' time finishing all five questions to analyze the effect of GistVis on their overall reading time.

For user perception, we employed a standardized test for workload, NASA-TLX~\cite{hart1988development}, to gauge users' workload under the two conditions. We used the Qualtrics online version created by \citet{castro2022examining} for our evaluation. Since user perception is highly personalized, we also collected qualitative data through a semi-structured interview. We asked participants questions regarding their perception of GistVis's impact on the reading experience, reading strategy, comments on the generated visualizations, and overall system verdict (see Supplementary Material for interview script).

Additionally, we used a short-form graph literacy test by \citet{okan2019using} to provide an anchor to validate participants' claims of their expertise in reading visualizations. We chose this short-form test over tests such as VLAT~\cite{lee2017vlat} or mini-VLAT~\cite{pandey2023minivlat} to prioritize pace over comprehensiveness to better control the time of our study. We used graph literacy as an additional variable and conducted regression analyses to investigate its impact on user performance. 

\begin{figure*}[tb]
  \centering
  \includegraphics[width=0.85\linewidth]{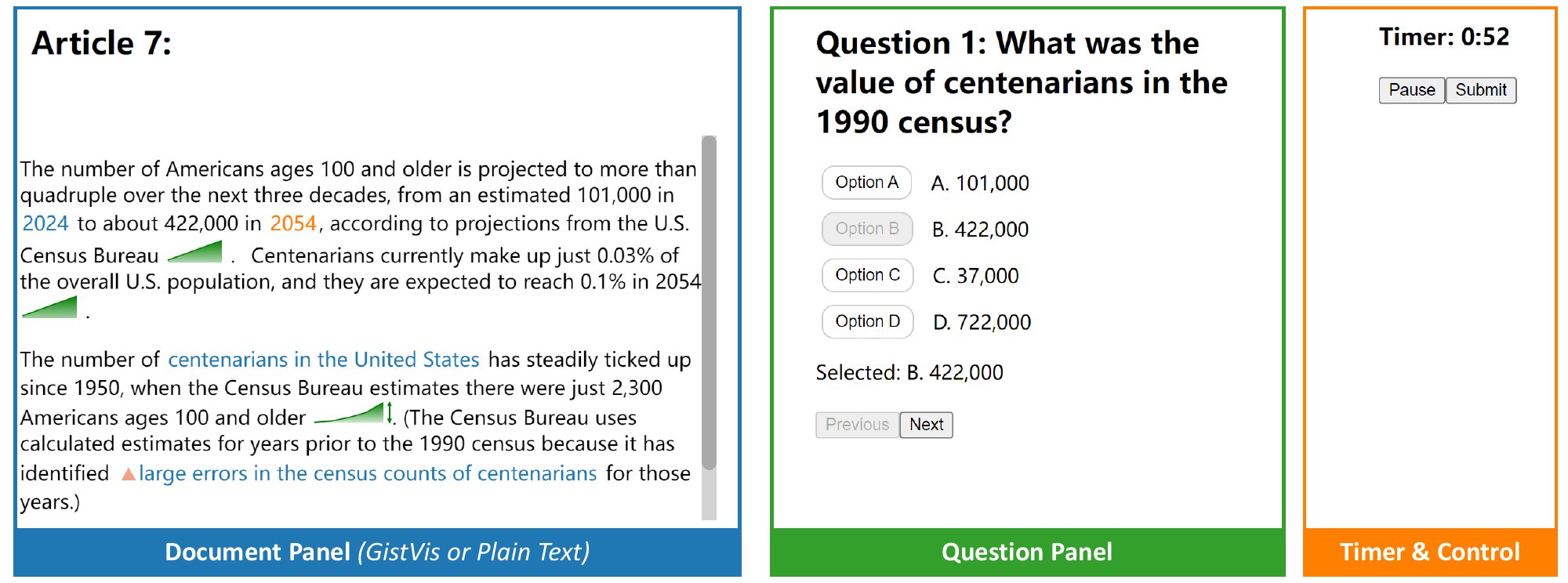}
  \caption{The interface we employed for our user study. The data-rich document is rendered in the \textbf{Document Panel} on the left. In the middle is the \textbf{Questions Panel}, where participants answer multiple-choice or summary questions. The right is the \textbf{Time and Control Panel}, where we record the finishing time for this passage when participants click on the submit button.}
  \label{fig:User-Study-Interface}
\end{figure*}

\subsection{Materials}
We chose six text articles from Pew Research, each from a different topic. All the articles could stand alone with text-only descriptions and have relevant visualizations removed from the original documents. We use GistVis to generate interactive word-scale visualizations for each article. Meanwhile, we used the exact specification but coerced all data fact types to no type to render a visualization-free version in Plain Text.

\begin{itemize}
    \item The \textsc{Age and Generations} article~\cite{us-centenarian} is about the growth of centenarians in the US over the next 30 years. The article comprises 646 words, 38 numbers, and 24 word-scale visualizations. Seven numbers and four word-scale visualizations are relevant to the task. 
    \item The \textsc{Science} article~\cite{us-k12-education} is about the perceptions and statistics for the United States kindergarten to 12th grade Science, Technology, Engineering, and Mathematics education. The article comprises 576 words, 22 numbers, and 24 word-scale visualizations. Seven numbers and six word-scale visualizations are relevant to the task. 
    \item The \textsc{Family and Relationships} article~\cite{age-of-us-couples} is about the narrowing age gap between U.S. husbands and wives over the past 20 years. The article comprises 580 words, 25 numbers, and 21 word-scale visualizations. Seven numbers and four word-scale visualizations are relevant to the task. 
    \item The \textsc{Internet and Technology} article~\cite{online-shopping-sales} is about trends in online shopping during the holiday season in the U.S. The article comprises 590 words, 24 numbers, and 21 word-scale visualizations. Four numbers and four word-scale visualizations are relevant to the task. 
    \item The \textsc{Race and Ethnicity} article~\cite{restaurants} is about the prevalence and distribution of Asian cuisine in U.S. restaurants. The article comprises 684 words, 34 numbers, and 27 word-scale visualizations. Seven numbers and four word-scale visualizations are relevant to the task. 
    \item The \textsc{Politics and Policy} article~\cite{black-voters-support} is about Black voters' preferences and political inclination for the 2024 election. The article comprises 659 words, 35 numbers, and 32 word-scale visualizations. Five numbers and five word-scale visualizations are relevant to the task. 
\end{itemize}

We chose these six articles because they are well-structured data documents and contain numerical data paired with text descriptions. They also cover various data fact types, making them ideal testbeds for our evaluation. Note that GistVis sometimes generated more visualizations than numbers. This is because GistVis generates visualizations for both explicit and implicit data insights based on the semantics of the description (for trend and extreme data types).

\begin{figure*}[tb]
  \centering
  \includegraphics[width=0.93\linewidth]{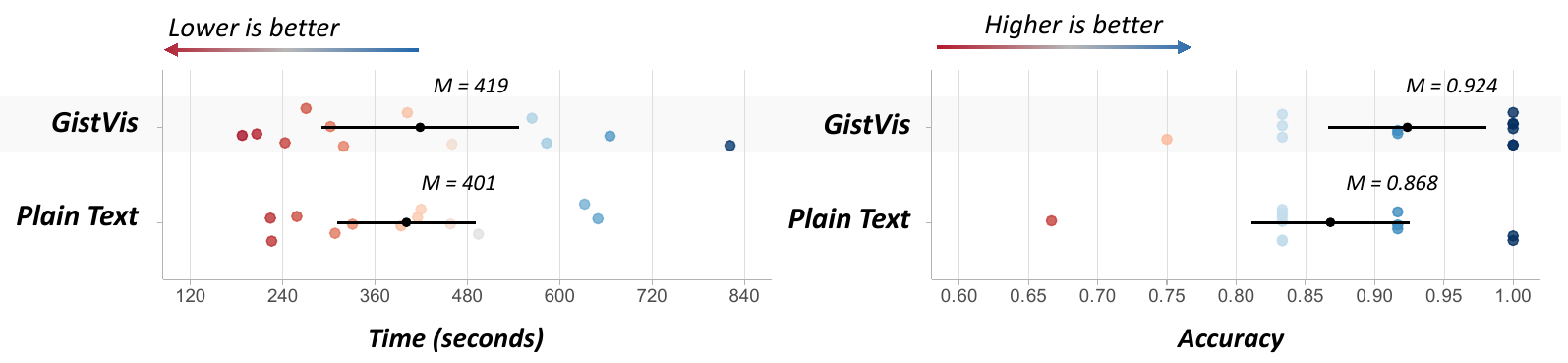}
  \caption{A comparison of user performance under GistVis and Plain Text over two utility metrics. Each scatter point in the graph represents one individual record from one participant. We map the values with a divergent color scheme from red (low) to blue (high). The black dot and the black line indicate the mean and 95\% confidence interval of the metric.}
  \label{fig:utility}
\end{figure*}

\begin{figure*}[tb]
  \centering
  \includegraphics[width=1\linewidth]{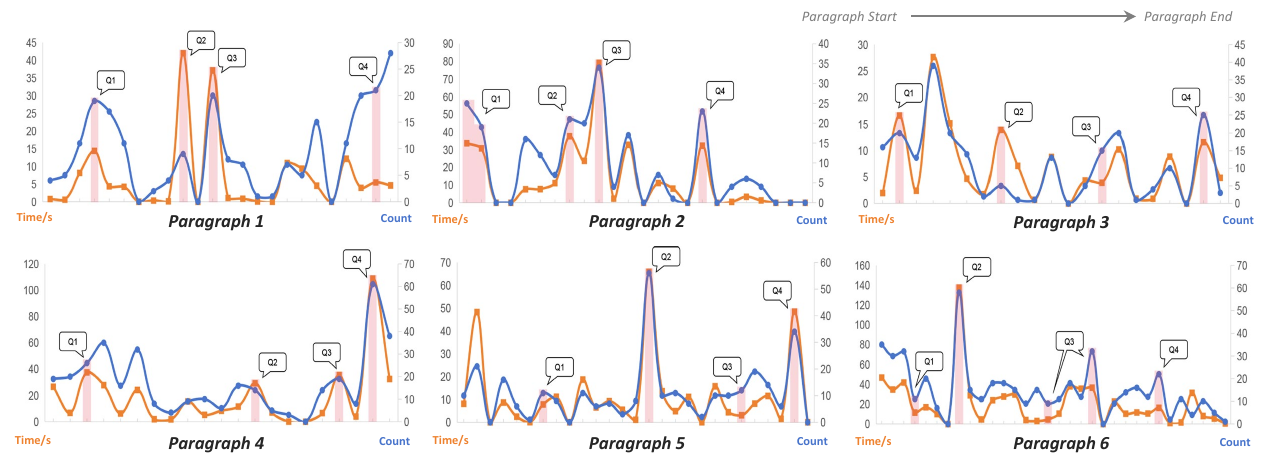}
  \caption{The aggregated interaction log from all 12 participants over six test passages. The x-axis represents the word-scale visualizations in the sequence of its occurrence. The yellow line indicates the time consumed on each word-scale visualization, whereas the blue line indicates the number of visits to the word-scale visualization. The transparent red bracket indicates the word-scale visualizations that contain the correct answers for our multiple-choice questions.}
  \label{fig:interation-log}
\end{figure*}

\subsection{Procedure and Tasks}
\paragraph{Introduction and Exploration}
Before the experiment, we conducted an onboarding process to familiarize participants with GistVis. We first provided a brief about the tasks of the experiment and then introduced the functionalities of GistVis to the participants. We allowed them to freely experiment with GistVis using a demo page that included a synthetic article not related to the text articles we used for the experiment.

\paragraph{Utility Tasks}
After the onboarding process, participants were asked to answer 30 questions from the above-mentioned six text articles, which contained 24 multiple-choice questions and six summary questions. Participants saw the Plain Text and GistVis conditions alternatively as they completed three articles for each condition. Researchers instructed participants to start with a specific text article under one condition during the experiment. Half of the participants started with the GistVis condition, while the rest started with the Plain Text condition. The sequence was also rotated every six participants. Researchers provided no intervention during the experiment except for technical support or clarifying the task requirement. For the last two articles, participants completed two rounds of the NASA-TLX test, one for each condition. We ran a timer on the interface for each text article participants go through. However, we did not impose time pressure on the tasks and allowed participants to finish them at their own pace.

\paragraph{Graph Literacy Test}
After finishing the 30 questions over six text articles, we asked participants to participate in a graph literacy test. We provide no time limit for this task.

\paragraph{Semi-structured Interview}
At the end of each session, we conducted a semi-structured interview to collect qualitative feedback on GistVis. We asked them about their preference between two conditions, their comment on the GistVis functionalities, reading strategies, the limitations of GistVis, and potential use cases. We also asked follow-up questions based on the participant's responses.

\subsection{Participants}
We recruited 12 participants (19 to 46 age range, M=22.8 years, SD=7.4 years, 8 self-identified as female, 4 as male) using the snowball sampling technique through social media. Most participants (9 undergraduate students, 2 master's students with a bachelor's degree, 1 graduate who is currently employed) were university students or graduates coming from various majors or occupations. On a 5-point scale, participants self-reported moderate expertise in visualization (M=2.3, SD=1.1) and reading data documents (M=2.9, SD=1.0). Additionally, because our experiment was conducted in English, we requested all non-native English speakers to self-report their English fluency. All participants who registered through our survey subsequently qualified our criterion to participate.

\subsection{Apparatus}
Out of 12 participants, 9 took part in the study remotely from their personal computers, with 3 participants taking part in person. Participants were permitted to control the researcher's computer (remotely through Tencent VooV-meeting's\footnote{\url{https://voovmeeting.com/}} remote control function and in person by directly using the researcher's computer) during the experiment. We implemented a user study interface (Fig.~\ref{fig:User-Study-Interface}) where participants would view and answer the question on the same interface. Under the participant's consent, we recorded participants' interaction with the website (e.g., answers to questions, movements, time) through our local backend. At the same time, we also screen-recorded the interaction process and audio-recorded participants' feedback.

\begin{figure*}[t]
  \centering
  \includegraphics[width=1\linewidth]{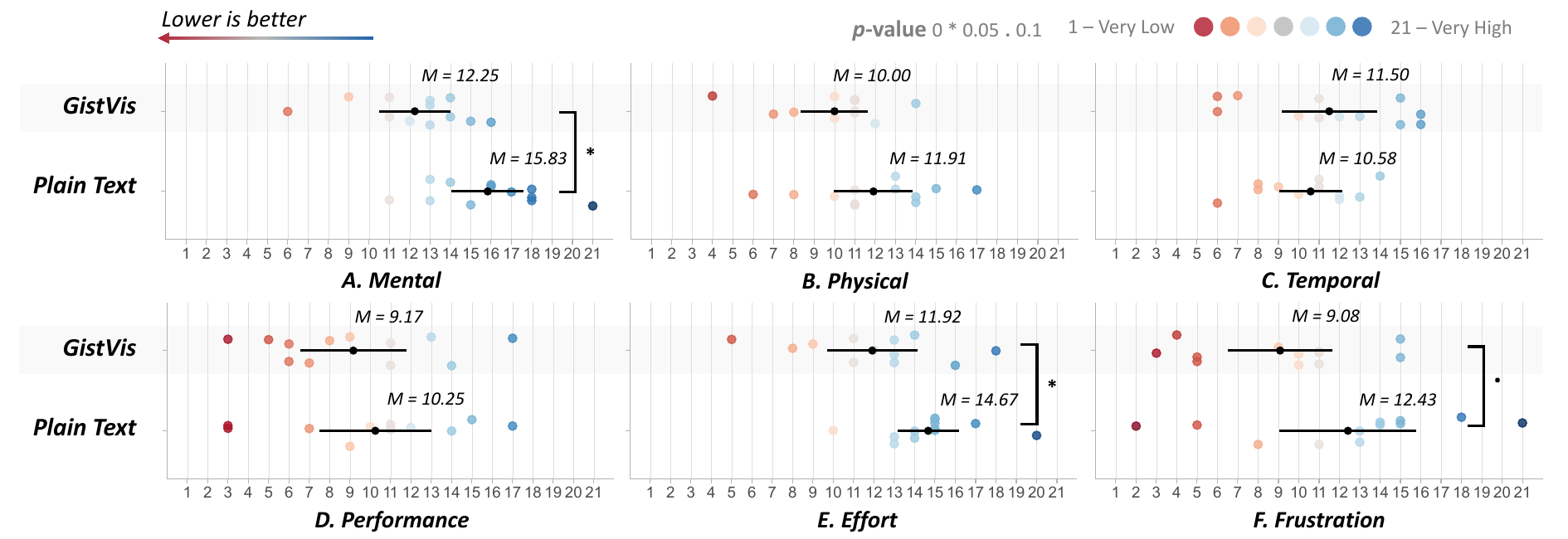}
  \caption{A comparison of perceived workload between GistVis and Plain Text over NASA-TLX metrics. Each scatter point in the graph represents one individual record from one participant. We map the values with a divergent color scheme from red (1-very low) to blue (21-very high). The black dot and the black line indicate the mean and 95\% confidence interval of the metric.}
  \label{fig:nasa-tlx}
\end{figure*}

\subsection{Results}
Since all participants experienced both conditions, we performed a paired t-test when values are normally distributed (time) and a Wilcoxon Signed-Rank test for other metrics (accuracy, workload). We analyzed our data using the statistical computing software R~\cite{r2013r} and reported corresponding statistical indicators (p-value, t-value for t-tests, and V-value for the Wilcoxon Signed-Rank test).

\subsubsection{Utility: Time}
Participants spent a similar amount of time answering the questions, as the paired test did not reveal a significant difference between the two conditions (p=0.536, t=0.639). Specifically, participants spent an average of 401 seconds (SD=175s) with the Plain Text condition and 419 seconds (SD=247s) with the GistVis condition (Fig.~\ref{fig:utility}). An additional regression analysis using linear models (R notation $\text{\textit{Time}} \sim \text{\textit{Condition}} + \text{\textit{GraphLiteracy}}$, referent Plain Text condition) revealed that participants' graph literacy would slightly reduce finishing time. However, the effect was minor and insignificant ($\beta$=-0.826, p=0.971).

\subsubsection{Utility: Accuracy}
Participants' accuracy was similar, as the Wilcoxon Signed-Rank test did not reveal a significant difference between the two conditions (p=0.182, V=14). However, visual analysis (Fig.~\ref{fig:utility}) indicated a trend towards better performance under the GistVis condition (M=0.924, SD=0.090) compared to Plain Text (M=0.868, SD=0.090). Notably, GistVis had a higher minimum accuracy (0.75) and a larger proportion of participants achieving a perfect score (6/12) versus Plain Text (2/12, minimum accuracy 0.67). We also ran a regression analysis to explore if graph literacy played a part in the accuracy (R notation $\text{\textit{Accuracy}} \sim \text{\textit{Condition}} + \text{\textit{GraphLiteracy}}$, referent Plain Text condition). Results indicated a positive effect for graph literacy on accuracy, yet the effect was small and insignificant ($\beta$=0.011, p=0.509).

\subsubsection{Utility: Interaction Log}
Visual analysis of the user interaction logs (Fig.~\ref{fig:interation-log}) revealed that participants benefited from the GistVis visualizations to complete the task. We observed in 81\% of the occasions, interaction (visit count or visit time) peaked around the word-scale visualization containing the correct answer for the question. Such evidence suggested that participants benefited from the automatically generated word-scale visualizations when tasked to retrieve data from data-rich documents.

On average, participants use interaction features most when answering questions related to the \textsc{Difference} data fact type, with the highest average dwell time (M=8.8 seconds) and hover count (M=5.2 times) per component. When addressing questions of \textsc{Proportion} data fact type, participants tend to take the time to understand the implication of the data insight with longer interaction time (M=4.7 seconds) with fewer visits to the component (M=3.1 times). Interaction is relatively minimal for questions related to other visualization components -- participants can quickly obtain information by viewing the charts without using the interactive feature. For instance, questions related to the \textsc{Trend} data fact type involve little interaction, with interaction time (M=3.5 seconds) and hover counts (M=3.4 times) on the lower side among all components. The remaining components exhibited even less user interaction than trend questions.

\subsubsection{Perception: Workload}
The Wilcoxon Signed-Rank test indicated that GistVis significantly reduced participants' mental demand (p=0.016, V=51.5) and effort (p=0.033, V=66.5) while revealing a marginal reduction in stress (p=0.059, V=46.1) compared to the Plain Text condition. We found no significant effect on physical, temporal, and performance. Meanwhile, a visual analysis revealed that compared to Plain Text, GistVis had a lower response (lower is better) in mental demand (M=12.25 versus M=15.83), physical demand (M=10.00 versus M=11.91), performance (M=9.17 versus M=10.25), effort (M=11.92 versus M=14.67) and frustration (M=9.08 versus M=12.43), while bearing a slightly higher temporal demand (M=11.50 versus M=10.58).

\subsection{Perception: Participant Feedback}
To seek additional user feedback, we analyzed the qualitative data collected through our semi-structured interview. Generally, participants found the visualization generated by GistVis satisfactory and useful, while we also identified room for improvements proposed by our participants. 

\paragraph{Highlighting entities and numbers.}
Eight out of twelve participants (P2, P3, P4, P5, P7, P8, P10, P11) mentioned that highlighting entities and numbers was helpful for their reading. The highlights enabled the efficient extraction of key information and facilitated the navigation of specific data while reading. P5 noted that the highlighted ``terms'' (entities) influenced his strategy of reading, as he utilized the highlighted items to locate key data-related terms:
\begin{quote}
\textit{``
For example, if there are references to years like 2023 or 2020 in the topic, I would first check if there are any visual indicators marking those years. Some terms might be directly highlighted in the article, so I would look for those terms and then examine the proportions or the data to make things easier.
    ''} - P5
\end{quote}
Meanwhile, while P10 claimed she retained her habit of reading, she also believes the highlighting helped her better understand the details in the documents:
\begin{quote}
\textit{``
If there are no highlights, I would read the entire text in detail, then outline the structure of the article and summarize the main content of each section. If there are highlights, my general strategy would be similar, but I might read (obtain) some of the details a bit more quickly.
    ''} - P10
\end{quote}

\paragraph{Participants perceive GistVis as an effective addition, especially for long data-rich documents.}
Seven participants found GistVis helpful for reading long texts (P4, P5, P6, P7, P8, P10, P12), especially when reading unfamiliar data-rich texts. P4, P6, and P8 believe that GistVis could increase document readability, especially for long data-rich documents, thereby reducing the mental load:
\begin{quote}
\textit{``
(Without visualization,) a long piece of text can be quite difficult to read. For instance, if you're submitting a data report to your advisor, they might be reluctant to read it. However, with such visual elements, they might be willing to spend a few valuable minutes reviewing it.
    ''} - P4
\end{quote}
\begin{quote}
\textit{``
(And) Having more labeled data makes it (data-rich documents) look much better. With these elements, reading a long data article becomes more engaging and novel. Yes, it makes it more likely that people will be willing to read it.
    ''} - P6
\end{quote}
P8 perceived the visualizations generated by GistVis, an effective proxy for data retrieval, which is useful for reading data-rich documents:
\begin{quote}
\textit{``
Without visualization, you might need to find the data yourself and perform calculations and comparisons on your own.
    ''} - P8
\end{quote}
P10 thought that GistVis would be helpful even for articles with pre-existing charts as it could strengthen the impression of the document and the understanding of the charts:
\begin{quote}
\textit{``
Or during the process of reading (an academic paper), you might need to jump back and forth between the text and the (original) charts. If you first read the text with visualizations, then look at the charts, it can reinforce your impressions.
    ''} - P10
\end{quote}

\paragraph{Rank, proportion, and trend are intuitive, but some visualization markings are hard to distinguish.}
Six participants (P3, P5, P6, P8, P10, P12) pointed out that the visualization we designed for rank, proportion, and trend is intuitive and easy to understand. However, participants identified room for improvement in visualizing some data fact types. For example, P4 and P5 suggested that using bar charts to represent both comparison and rank makes it visually challenging to discern the different data insights without interacting with them: 
\begin{quote}
\textit{``
For example, when (GistVis) visualizes differences, it uses bar charts, and when listing each data point, it also uses bar charts. Sometimes, just looking at these small charts, you can't immediately tell whether they represent the difference or the individual values (rank).
    ''} - P4
\end{quote}
P4 also suggested using a coherent color mapping for the same entity occurring multiple times:
\begin{quote}
\textit{``
(Highlighting of existing entities with) new colors not previously used can be confusing. It would be better if the colors were consistent throughout.
    ''} - P4
\end{quote}

\section{Discussion and Future Work}
Our work explored using the capabilities of LLMs to generate interactive word-scale in situ visualizations from data-rich documents. In this section, we describe the limitations of GistVis and envision future research opportunities.

\paragraph{Understanding the Effectiveness of Word-scale Visualizations with Finer Granularity}
Our findings suggested that GistVis is effective in helping users understand data-rich documents following a document-centric strategy. However, the design of our evaluation limited us in obtaining finer-grained information about which part of the word-scale visualization made them particularly effective. Specifically, our qualitative data from the semi-structured interviews and the interaction log revealed interesting patterns that require further investigation. For example, we did not find strong evidence of the effectiveness of the drill-down operation with the tooltip, as no participant mentioned using those as part of their strategy. Meanwhile, our interaction log revealed that users interact with few word-scale visualizations pertinent to the question. However, exactly what they gazed at before their decisions remained unclear. Future work could use eye-tracking~\cite{huth2024eye} to explore the gaze pattern of users to understand their underlying strategy processing GistVis augmented documents.

\paragraph{Lifting the Constraints of GistVis}
We built GistVis around a series of rather stringent constraints. Initially, we defined several constraints based on the results of our formative study on data-rich documents before implementation (Sec.~\ref{subsec:formative-constriants-implications}). While this practice allowed us to conduct a proof-of-concept study, we inevitably ignored several conditions and cases in the wild that might be equally important to the conditions we supported. Additionally, we incorporated further constraints through the design of our computational pipeline. For instance, we assumed that the same data insight should be located in proximity to each other and within the same paragraph. However, there are instances where the same data insight may span across multiple paragraphs or even the entire passage. While we contend that our system supports the majority of scenarios appropriate for document-centric analysis with word-scale visualization, especially professionally written data documents (e.g., \cite{us-k12-education}), future research should aim to flip these constraints and expand GistVis to encompass a wider range of conditions.

\paragraph{Collect a Comprehensive Dataset of Data-rich Documents for Analysis and Training}
Before designing GistVis, we collected a corpus containing 44 data-rich documents to conduct a formative analysis of data-rich documents. We also utilized this annotated dataset to quantitatively evaluate the performance of GistVis. However, our corpus is far from comprehensive because we only included the genre of data journalism. Other data-rich documents, such as business reports and scientific papers, might have different narrative features to data journalism and should be thoroughly analyzed to build a firm foundation to expand the capability of our GistVis pipeline. Moreover, a larger, more reliable, and more comprehensive dataset could also benefit the design and evaluation of better algorithmic approaches. For example, future work could implement better in-context learning strategies, such as Active Learning inspired strategies (e.g., In-context Sampling~\cite{yao2024more}), to improve the few-shot capabilities of LLMs. We advocate future research to propose methods to collect and generate large-scale datasets for visualization~\cite{wu2022ai4vis} so that visualization researchers can better understand the narrative features of word-scale visualizations and improve automatic algorithms for word-scale visualization generation.

\paragraph{Expand the Design Space of Word-scale Visualizations}
Our current implementation of GistVis only contains three different chart types with 14 visualization variants, including icons. More chart types and different variants tailored for unique narrative features should be implemented to further improve the usability of the pipeline. For example, in some cases, pie charts might better convey proportional insight than the horizontal stacked bar chart we applied in this work. Future work could include pie charts in our design space while working on heuristics or algorithmic approaches to guide GistVis in selecting the better representation based on the text context. In addition to charts and icons, future work could also discuss using novel typefaces~\cite{nacenta2012fatfonts} or tiny graphics~\cite{zhao2020iconate}. We argue that the modular design of GistVis would enable us to easily expand the search space of word-scale visualizations.

\paragraph{Improve Interactivity}
Although GistVis supports basic interactivity, the scope of interaction is limited within a unit segment. One typical instance that reflects this limitation is when the same entity exists across unit segments: our current implementation would likely label the same entity with different colors. Such practice would potentially lead to confusion about the document's content. Moreover, there might be multiple word-scale visualizations related to the same entity, and our current approach could not synchronously show all the entity's related data. Future work should seek to employ universal control over the interactive components, including a color mapping system for entity label consistency and synchronous interaction for the same entities. 

Additionally, although we focus on the document-centric analysis of data-rich documents, users might benefit from being able to deploy both document-centric and visualization-centric analysis. For example, for data insights that span multiple paragraphs, interaction techniques such as view manipulation~\cite{heer2012interactive} could reorganize word-scale visualizations to support the transition between document-centric analysis and visualization-centric analysis~\cite{goffin2020Interaction}. Additionally, some data-rich documents may already have figure-size visualizations that support document-centric analysis. Although one participant (P10) believes GistVis could also be effective even when figure-size visualizations exist, we need more evidence to ascertain its effectiveness and determine a strategy to combine both forms of augmentation. Thus, future work should also explore interaction techniques to synchronize text, word-scale visualizations, and figure-size visualizations to enable users to benefit from document-centric and visualization-centric analysis.

\paragraph{Integrating GistVis into Existing Workflow}
Participants expressed a keen interest in using GistVis if it were integrated into their existing workflow. While this work validates the potential of using automatically generated word-scale visualizations from GistVis, extensive engineering efforts are required before it can be fully supported in real-world workflows. GistVis holds the potential to be applied in a wide array of use cases, such as functioning as a plug-in for document readers or enhancing the capabilities for visual analysis tools~\cite{ava-ntv-2024}.

\section{Conclusion}
In this paper, we proposed an automatic pipeline, GistVis, for generating word-scale visualization using LLMs. We informed our design with a formative study across 44 data-rich documents. We designed GistVis modularly to support plug-and-play property for expansion while we steer LLMs with visualization knowledge to generate word-scale visualizations to support document-centric analysis. Our technical evaluation and user study reveal that GistVis could generate satisfactory word-scale visualizations that could reduce users' workload reading data-rich documents. We discuss the limitations of our current approach and outline future directions. We believe that GistVis is a timely contribution to inspire further study from the visualization community about using automatic methods, especially LLMs, to augment data-rich documents.

\begin{acks}
This work was supported in part by the Shanghai Student Innovation Training Program \#S202410247252. We want to thank all participants for taking part in our study. Additionally, we extend our thanks to the anonymous reviewers for their constructive feedback.
\end{acks}

\balance

\bibliographystyle{ACM-Reference-Format}
\bibliography{bibs/auto-docuvis, bibs/web-link}

\end{document}